\documentclass{PoS}
\usepackage{style}
\graphicspath{{figs/}}

\title{Semileptonic $B_c$ decays from full lattice QCD}

\ShortTitle{Semileptonic $B_c$ decays from full lattice QCD}

\author{
\speaker{Andrew Lytle},
Brian Colquhoun,
Christine Davies,
Jonna Koponen
\\
SUPA, School of Physics and Astronomy, 
University of Glasgow, Glasgow, G12 8QQ, UK\\
E-mail: \email{andrew.lytle@glasgow.ac.uk}
}

\author{
Craig McNeile\\
Centre for Mathematical Sciences,
Plymouth University, Plymouth, PL4 8AA, UK
}

\author{HPQCD Collaboration\thanks{URL: http://www.physics.gla.ac.uk/HPQCD}}

\abstract{
We present first lattice QCD results for semileptonic form factors
of the decays
$B_c \rightarrow \eta_c \, l\nu$ and $B_c \rightarrow J/\psi \, l\nu$
over the full $q^2$ range, using both improved non-relativistic QCD (NRQCD)
and fully relativistic (HISQ) formalisms.
These can be viewed as prototype calculations
for pseudoscalar to pseudoscalar and pseudoscalar to vector decays involving
a $b \rightarrow c$ transition. In particular we can use information from
the relativistic computations to fix the NRQCD current normalisations,
which can then be used in improved computations of 
decays such as $B \rightarrow D \, l \nu$ and $B \rightarrow D^* \, l \nu$.
}

\FullConference{16th International Conference on B-Physics at Frontier Machines\\
		2-6 May 2016\\
		Marseille, France}

\begin{document}

\section{Introduction}
Semileptonic decays of $B$-mesons provide the main inputs for exclusive
determinations of $|V_{ub}|$ and $|V_{cb}|$
(a status update of lattice QCD's impact on these quantities was
presented at this conference~\cite{VandeWater}). The precise determination of
$|V_{cb}|$ requires precision in both theoretical computation and measurement
of $b \rightarrow c$ processes. Treatment of $c$ and especially $b$ quarks
is a challenge
for lattice simulations due to lattice artifacts which grow as $(am_q)^n$ where
$a$ is the lattice spacing and $m_q$ is a quark mass. We have two complementary
approaches to the treatment of $b$ quarks: using a highly improved relativistic
action at small lattice spacings to simulate masses approaching $m_b$,
and working directly at $m_b$ with an improved non-relativistic (NRQCD)
effective theory formalism.

\section*{Methodology}
We use a highly improved staggered quark (HISQ)
action~\cite{Follana:2006rc} which systematically
removes lattice artifacts, allowing for simulation of charm
quarks with small discretisation effects. We can even simulate
quarks with mass significantly larger than $m_c$, especially on the
ensembles with very fine lattice spacings 
of $a \approx 0.045$ fm~\cite{McNeile:2012qf}.
This motivates one of our approaches for doing
$b$-physics.
By working in a regime $am_h \lesssim0.8$, say,
but on finer and finer lattice spacings,
we calculate the physics of interest over a range in $m_h$ and then
extrapolate that data to $m_b$.

We are also able to work directly at the $b$ mass, without the need for
extrapolation, using an improved non-relativistic 
(NRQCD) formalism~\cite{Lepage:1992tx}.
This approach is complementary to the fully relativistic approach described
above.
The NRQCD Hamiltonian is expressed as an expansion in the
velocity of the heavy quark.
In addition the current operators have a relativistic expansion,
e.g.\ the temporal axial-vector current
\begin{equation*}
A_0^{\text{nrqcd}} = (1+\a_s z_0^{(0)})
\[A_0^{(0)} + (1+\a_s z_0^{(1)})A_0^{(1)} + \a_s z_0^{(2)} A_0^{(2)} \] 
+ \dotsc \,,
\end{equation*}
where $A_0^{(1)}$, $A_0^{(2)}$, $\dotsc$ are higher order current corrections,
with matrix elements proportional to $1/m_b$.
The matching to the continuum current above is only known
in QCD perturbation theory to $\O(\a_s)$,
and so has a systematic uncertainty from missing $\a_s^2$ terms.
One output of the present work will be to improve the normalisation
of the currents using the fully relativistic calculation where the
normalisation is simpler and nonperturbative.

We use gauge ensembles generated by the MILC collaboration, which include the
effects of $u/d$, $s$, and $c$ quarks in the sea, and with lattice spacings
of $a \approx 0.09, 0.06$, and $0.045$ fm. Although we have ensembles with
physical $u/d$ quark mass, all results presented here use $m_s/m_{u/d} = 5$,
i.e.\ unphysically heavy pions.

\subsection*{Obtaining the form factors}
In both formalisms we compute the matrix element of the $V-A$ operator
between the states of interest. We work in the frame where the $B_c$
is at rest. The matrix elements are parametrised in terms of form factors
which are functions of $q^2$, where $q = P-p$ is the difference
in four-momentum between the $B_c$ and outgoing particle.
The kinematic endpoint $q^2_{\text{max}}$ is
where the outgoing hadron is at rest, whereas the energy of the outgoing
hadron is a maximum at $q^2=0$.

For the $B_c \rightarrow \eta_c$ decay there are two form factors
to determine, $f_+$ and $f_0$.
\begin{equation*} \label{ps-ps}
\bra{\eta_c(p)} V^{\mu} \ket{B_c(P)} =
f_{+}(q^2) \[P^{\mu} + p^{\mu} - \frac{M^2 - m^2}{q^2} q^{\mu} \] +
f_{0} (q^2) \, \frac{M^2 - m^2}{q^2} \, q^{\mu}
\end{equation*}
For the relativistic
case we can determine $f_0$ using a scalar current, which is absolutely
normalised.
\begin{equation*}
\bra{\eta_c(p)} S \ket{B_c(P)} = \frac{M^2-m^2}{m_{b0}-m_{c0}}f_0(q^2)
\end{equation*}

There are five form factors to determine for the
$B_c \rightarrow J/\psi$ decay, one from the vector current and four
from the axial-vector current (three of which are independent).
\begin{multline*}\label{eq:ps-v}
\bra{J/\psi(p,\ve)} V^{\mu} - A^{\mu} \ket{B_c(P)} = 
\frac{2i \e^{\mu \nu \rho \sigma}}{M + m}
\ve^{*}_{\nu} p_{\rho} P_{\sigma} \, V(q^2) -
(M + m) \ve^{*\mu} \, A_1(q^2) + \\
\frac{\ve^{*} \cdot q}{M + m} \(p + P\)^{\mu} \, A_2(q^2) +
2 m \frac{\ve^{*} \cdot q}{q^2} q^{\mu} \, A_3(q^2) -
2 m \frac{\ve^{*} \cdot q}{q^2} q^{\mu} \, A_0(q^2)
\end{multline*}

\section*{Results}
\begin{figure}
\includegraphics[width=0.4\textwidth]{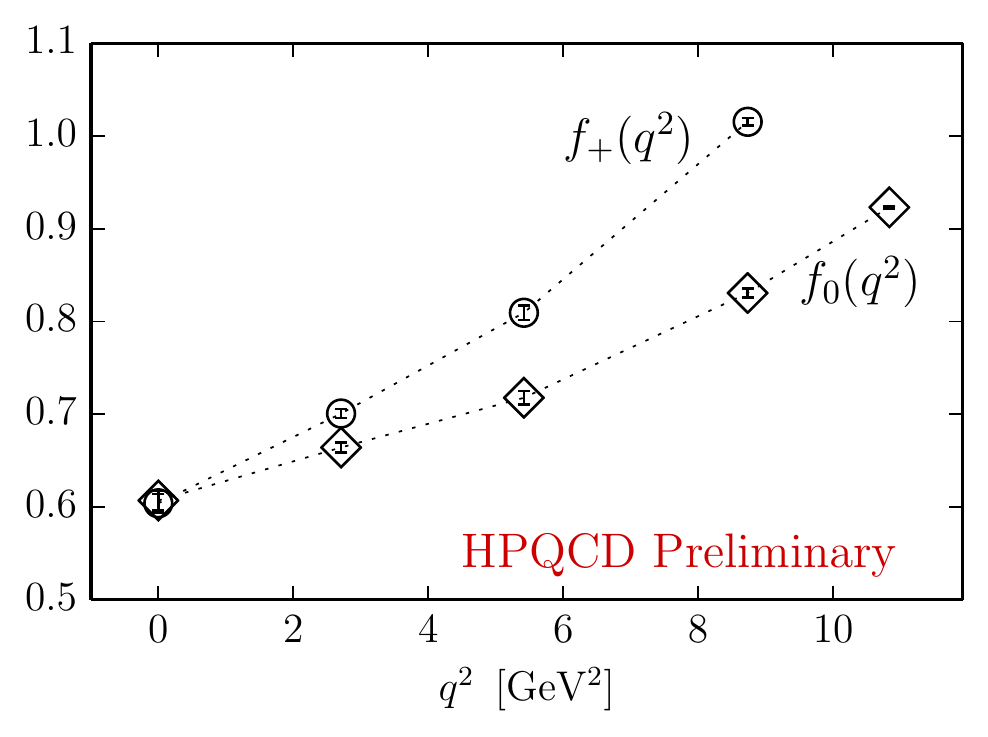}
\includegraphics[width=0.6\textwidth]{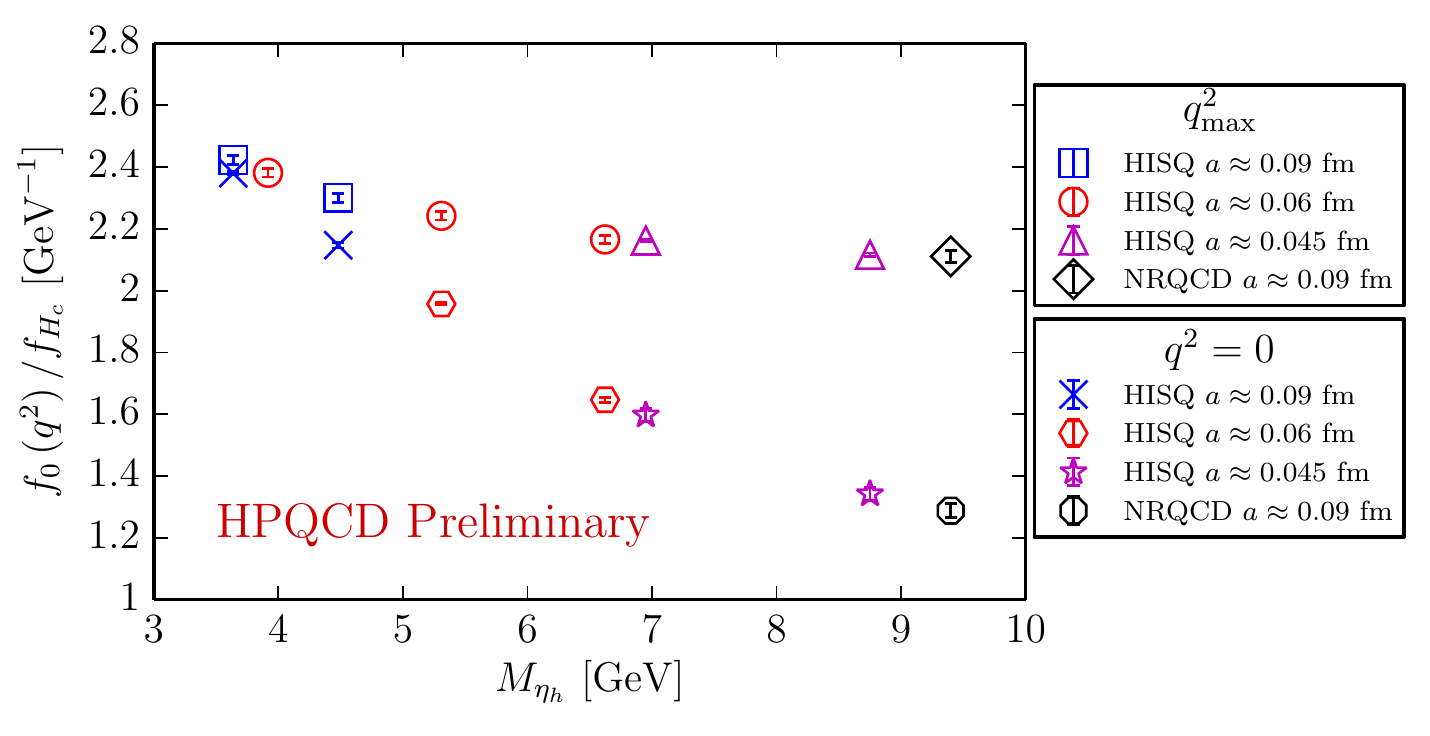}
\caption{
(Left)
Results for $B_c \rightarrow \eta_c$ form factors $f_0$ and $f_+$
from lattice NRQCD, determined over the full $q^2$ range.
(Right)
Extrapolation results in heavy quark mass $m_h$
for $f_0(q^2_{\text{max}})/f_{H_c}$ and $f_0(0)/f_{H_c}$
using the fully relativistic (HISQ) formalism. The rightmost
points are the corresponding NRQCD results with physical $b$ mass.
\label{f+-f0}}
\end{figure}

Figure~\ref{f+-f0} (left) shows our results for the $B_c \rightarrow \eta_c$
form factors $f_+(q^2)$ and $f_0(q^2)$ computed using improved NRQCD
on the $a \approx 0.09$ fm ensemble.

In Figure~\ref{f+-f0} (right) we show results for
$f_0(q^2_\text{max})/f_{H_c}$ and $f_0(0)/f_{H_c}$ using the relativistic
formalism on ensembles with lattice spacings
of $a \approx 0.09, 0.06$, and $0.045$ fm. For each ensemble we
use valence masses $m_h$ such that $a m_h \leq 0.8$, which correspond to
heavier physical masses as we go to smaller lattice spacings.
As $m_h$ approaches $m_b$ the results join smoothly with results computed
using NRQCD (rightmost points).

\begin{figure}
\centering
\includegraphics[width=0.5\textwidth]{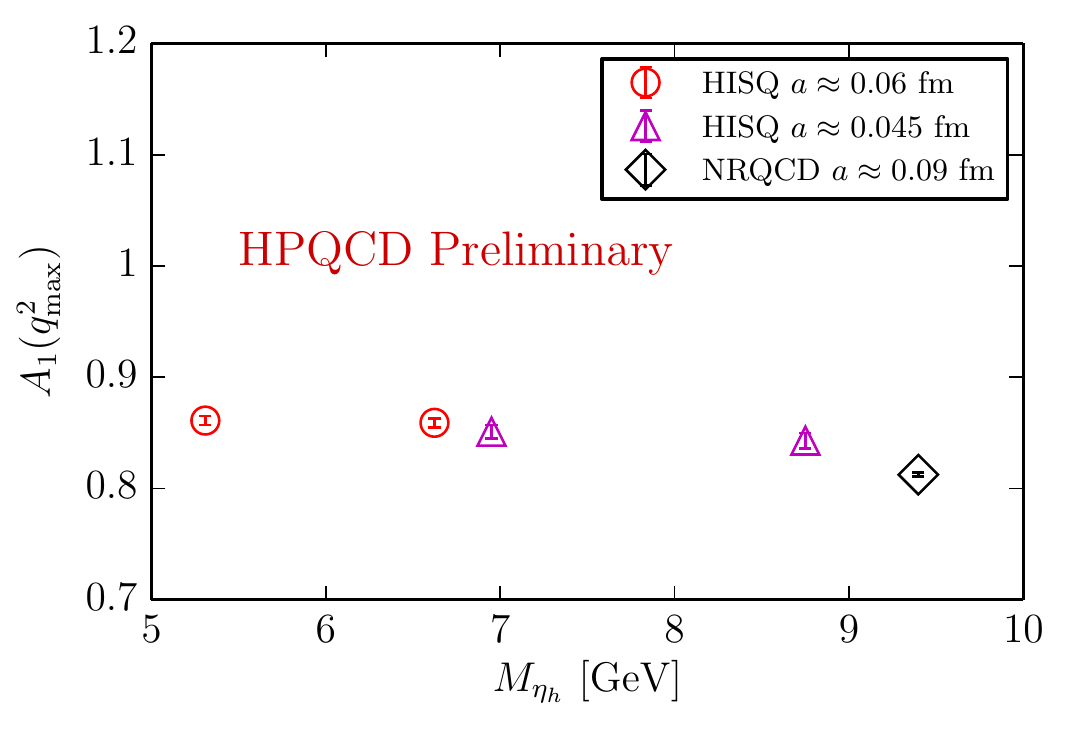}
\caption{
Extrapolation in heavy quark mass $m_h$ for the 
$B_c \rightarrow J/\psi$ form factor
$A_1(q^2_{\text{max}})$ using the relativistic HISQ formalism.
The rightmost point is the NRQCD result at physical $b$ mass.
\label{A1-q2}}
\end{figure}

Figure~\ref{A1-q2} is another extrapolation plot, this time showing
the $B_c \rightarrow J/\psi$ form factor $A_1(q^2_\text{max})$.
This is the only form factor that contributes to the
decay rate at zero recoil. Furthermore Luke's theorem tells us
the $1/m_b$ current corrections vanish there
so that the comparison between NRQCD and relativistic data is
particularly simple.

\section*{Conclusions}
We have presented results for the $B_c$ semileptonic decays to $\eta_c$ and
$J/\psi$ using two complementary approaches.

\begin{itemize}
\item The $B_c \rightarrow \eta_c$ results provide proof-of-principle
for our strategy; we are able to control the calculation over the full
$q^2$ range and find good agreement between the NRQCD and fully relativistic
approaches.

\item Our first results for the $B_c \rightarrow J/\psi$ decay also appear
promising. The full lattice calculation of this decay will allow
the extraction of $|V_{cb}|$ if the decay is measured at LHCb.

\item The NRQCD $b \rightarrow c$ currents also mediate the decays
$B \rightarrow D$ and $B \rightarrow D^*$. Using information from
the relativistic calculation we will improve the normalisations of the currents,
which will lead to improvements in theoretical precision for these decays.
\end{itemize}

\section*{Acknowledgements}
AL would like to thank the organisers for an enjoyable conference, and 
Gagan Mohanty, Greg Ciezarek, Barbara Sciascia, Lucio Anderlini,
and Jorge Martin Camalich for 
useful discussions. This work was performed on the Darwin supercomputer,
part of STFC's DiRAC facility.

\end{document}